\documentclass[aps,prl,amsmath,amssymb,superscriptaddress,showkeys,showpacs,twocolumn,floatfix]{revtex4-1}

\usepackage{amsmath,amssymb,mathtools}	
\usepackage{graphicx,color}
\usepackage{float}
\usepackage{appendix}
\usepackage{hyperref}
\usepackage{verbatim}
\usepackage{comment}
\usepackage{subcaption}
\usepackage{bm}
\usepackage{xcolor}
\usepackage[justification=raggedright,singlelinecheck=false]{caption}

\usepackage[utf8]{inputenc}
\usepackage[english]{babel}
\usepackage{booktabs}
\usepackage[normalem]{ulem}

\begin{document}
\title{Testing Nambu-Goto approximation of cosmic string by lattice field simulations}

\author{Zizhuo Zhao}
\affiliation{
~International Centre for Theoretical Physics Asia-Pacific (ICTP-AP), University of Chinese Academy of Sciences (UCAS), Beijing, China.
}
\affiliation{
~Taiji Laboratory for Gravitational Wave Universe (Beijing/Hangzhou), University of Chinese Academy of Sciences (UCAS), Beijing, China.}

\author{Ligong Bian\thanks{Corresponding Author.}}
\email{lgbycl@cqu.edu.cn}
\affiliation{~Department of Physics and Chongqing Key Laboratory for Strongly Coupled Physics, Chongqing University, Chongqing 401331, P. R. China}

\author{Jing Shu\thanks{Corresponding Author.}}
\email{jshu@pku.edu.cn}
\affiliation{~School of Physics and State Key Laboratory of Nuclear Physics and Technology, Peking University, Beijing 100871, China}
\affiliation{
~Center for High Energy Physics, Peking University, Beijing 100871, China}
\affiliation{
~Beijing Laser Acceleration Innovation Center, Huairou, Beijing, 101400, China}

\begin{abstract}
The precise calculation of gravitational wave (GW) from cosmic string networks is of significant theoretical and experimental interest. The Nambu--Goto (NG) approximation has long been employed to calculate GW emission from such networks; however, its validity has never been systematically verified. We perform large-scale zero-temperature Abelian-Higgs lattice simulations under different gauge couplings, and compare them with NG predictions. We find excellent agreement in the power-law region for near-global strings but strong deviation for strongly coupled local strings with $m_v/m_s \sim 1$, quantitatively establishing the breakdown of the NG approximation. Additionally, we confirm that particle emission significantly dominates the energy loss of the string network, with the ratio of GW energy to particle energy approximately $10^{-3}$ to $10^{-2}$ for both near-global and local string scenarios.

\end{abstract}	

\maketitle

\noindent{\it \bfseries Introduction.}
Cosmic strings, as one-dimensional topological defects, form when a U(1) symmetry is spontaneously broken~\cite{TWB_Kibble_1976}, and are promising detection targets for pulsar timing arrays \cite{PTA_1,PTA_2,PTA_3,PTA_4,PTA_5}, LIGO-Virgo~\cite{LIGO-Virgo_1, LIGO-Virgo_2}, and the Laser Interferometer Space Antenna (LISA) \cite{LISA_1}. Currently, most predictions of GWs from cosmic strings rely on the Nambu-Goto (NG) approximation, which, however, has never been systematically verified.  The dynamics of infinitely thin strings can be effectively described by the NG action when the string width can be ignored. In such a situation, intersections among infinite strings inevitably cause segment exchanges and loop formation \cite{texbook_cosmic_string}, driving the network towards a scaling regime. These loops oscillate and emit gravitational wave (GW) bursts from structures known as cusps and kinks \cite{cusp_kink_1, cusp_kink_2}, and the superposition of numerous uncorrelated GW bursts generates a stochastic gravitational wave background (SGWB). 

The GW spectrum depends on the number density distribution of string loops, characterized by the BOS model~\cite{PhysRevD.89.023512} and the LRS model~\cite{LRS_model} based on simulations of NG strings. These simulations do not require evolving fields on a lattice, but instead directly simulate the strings themselves~\cite{PhysRevD.72.063514,PhysRevD.75.063521,PhysRevD.83.083514}\footnote{ The energy of a long string network undergoes a redshift and is transferred to loops, which is effectively described by the velocity-dependent one-scale (VOS) model~\cite{texbook_cosmic_string}.}, and therefore lose the direct relation with the underlying field dynamics of symmetry breaking. In addition, lattice field simulations show that oscillating string loops can emit particles, such as axion-like particles from global strings and dark photons from local strings, both recognized as compelling dark matter (DM) candidates (see Ref.\cite{axion_DM_1, axion_DM_2, axion_DM_3} for axion-like particles and Refs.\cite{dark_photon_DM_1, dark_photon_DM_2} for dark photons).

In this work, we perform large-scale zero-temperature Abelian-Higgs lattice simulations under different gauge couplings. Our simulation is based on the entire string network rather than individual loops in previous studies~\cite{2001.01030, 2308.08456v2,Hindmarsh_2017, 1903.05102, 2103.16248, 2408.02364}. We directly compare the GW spectra obtained from NG and lattice simulations to test the validity of the NG approximation because lattice simulations provide a first-principles approach to capturing the finite string width and particle radiation.


Our results indicate that the validity of the NG approximation is controlled by whether the dominant GW-emitting modes remain well separated from the string-width scale. In particular, the approximation breaks down in the strongly coupled local case, where these modes probe the microscopic string core. In addition, we extract the universal transfer function of the GW spectrum directly from lattice data, providing a bridge to analytical extrapolations over cosmological times. For both the near-global and local scenarios, the networks lose most of their energy via particle emission.

\noindent{\it \bfseries The simulation setup.}
We consider an Abelian-Higgs model described by the Lagrangian:
\begin{equation}\label{eq:Lagrangian}
    \mathcal{L}=(D_{\mu}\Phi)^{\dagger}(D^{\mu}\Phi)-\frac{1}{4}F_{\mu\nu}F^{\mu\nu}-V(\Phi),
\end{equation}
where $\Phi$ is a complex scalar field, $D_{\mu}=\partial_{\mu}-ieA_{\mu}$ is the covariant derivative associated with the gauge field $A_{\mu}$, and $F_{\mu\nu}=\partial_{\mu}A_{\nu}-\partial_{\nu}A_{\mu}$ is the gauge field strength tensor. The potential is given by \cite{texbook_cosmic_string}:
\begin{equation}\label{eq:potential}
    V(\Phi)=\frac{\lambda}{4}(|\Phi|^2-v^2)^2,
\end{equation}
where $\lambda$ is the scalar field self-coupling constant and $v$ denotes its vacuum expectation value (VEV).  In the zero-temperature simulations considered in this work, the U(1) symmetry is already spontaneously broken, and the string network is generated by spatial variations of the initial field configuration. After symmetry breakdown, the scalar and gauge fields acquire masses $m_s=\sqrt{\lambda}v$ and $m_v=\sqrt{2}ev$, respectively, and the ratio $\beta=m_v^2/m_s^2=2e^2/\lambda$ significantly influences the properties of the string network. The string width is constrained by the Compton wavelengths of fields \cite{texbook_cosmic_string}, which is expressed as:
\begin{equation}\label{eq:string_width}
    r_s\sim m_s^{-1}=\Big(\sqrt{\lambda}v\Big)^{-1}, \quad r_v\sim m_v^{-1}=\Big(\sqrt{2}ev\Big)^{-1}\;.
\end{equation}
For numerical accuracy, the lattice spacing in simulations must be smaller than these characteristic lengths. As the physical grid interval scales proportionally with the scale factor $a(t)$, this imposes an upper bound on the feasible simulation duration.

In the subsequent simulations, we assume an FLRW metric as the background spacetime and use conformal time $\tau$ as the temporal coordinate. The scale factor is prescribed by a radiation-dominated background rather than being solved dynamically from the lattice fields; therefore, $a(\tau)\propto \tau$ in conformal time. Throughout this Letter, we adopt the temporal gauge $A_0=0$. Following Ref.~\cite{Cosmolattice, 2006.15122}, the leap-frog algorithm is employed to numerically calculate the equations of motion (EOM)\cite{2212.13573}:
\begin{align}\label{eq:EOMs}
    \Phi^{\prime\prime}+2\mathcal{H}\Phi^{\prime}-D_iD_i\Phi+a^2\frac{\partial V}{\partial\Phi^{\dagger}}&=0\;,  \\
    E_i^{\prime}+\partial_jF_{ij}-2ea^2\textnormal{Im}(\Phi^{\dagger}D_i\Phi)&=0\;, \\
    \partial_iE_i-2ea^2\textnormal{Im}(\Phi^{\dagger}\Phi^{\prime})&=0\;
\end{align}
where $\Phi^{\prime}=\frac{\partial \Phi}{\partial \tau}$ and $\mathcal{H}=\frac{a^{\prime}}{a}$ denotes the conformal Hubble parameter. Instead of evolving the gauge field $A_{\mu}$ directly, we evolve the "electric field" $E_i=F_{0i}$. The third equation is the Gauss constraint, arising naturally from gauge invariance. The code is available at \footnote{\href{https://github.com/ZizhuoZhao23/NG_test_simulation_code}{Code available at GitHub: \texttt{github.com/ZizhuoZhao23/NG\_test\_simulation\_code}}}. The quantum fluctuation spectrum is used to initialize the scalar field and its conjugate momenta, see the {\it Appendix} for details. In addition, we adopt the procedure outlined in \cite{2409.16124} to initialize the conjugate momenta of the gauge field $A_i^{\prime}$, ensuring that the Gauss constraint is satisfied, and set the initial gauge field to zero. 
 
The GW is generated through oscillations and intersections of cosmic strings. In lattice simulations, the evolution of the metric perturbation $h_{ij}$ is governed by
\begin{equation}\label{eq:eom_of_hij}
    h^{\prime\prime}_{ij}-\nabla^2 h_{i j}+2\mathcal{H}{h}^{\prime}_{ij}=16 \pi G T_{ij}^{\mathrm{TT}},
\end{equation}
where $T^{\mathrm{TT}}_{ij}$ is the transverse-traceless component of the energy-momentum tensor, defined as \cite{2408.02364}:
\begin{equation}\label{eq:energy-mumentum_tensor}
    T_{ij}^{\mathrm{TT}}=2\textnormal{Re}[(D_{i}\Phi)(D_j\Phi)^{\dagger}]+F_{i\alpha}F_{\beta j}g^{\alpha\beta},
\end{equation}
with $g^{\mu\nu}$ being the FLRW metric. The dimensionless GW spectrum is then calculated by:
\begin{equation}\label{eq:GW_spectrum_lattice}
    \Omega_{\text{gw}}(k)=\frac{1}{\rho_c}\frac{\partial \rho_{\text{gw}}(k)}{\partial \ln{k}}=\frac{1}{24\pi^2 V} \frac{k^3}{\mathcal{H}^2}\sum_{ij}|h_{ij}^{\prime}(k,\tau)|^2,
\end{equation}
where $h_{ij}(k,\tau)$ denotes the Fourier transform of $h_{ij}(x)$, $V$ is the comoving volume of the whole lattice and $\rho_c=3/(8\pi G)H^2$ is the critical energy density. When calculating the GW spectrum from the lattice simulation, we use the Runge-Kutta method to numerically solve Eq.\eqref{eq:eom_of_hij}, and the calculation of the GW spectrum within the lattice simulation is performed using the public code {\it pystella}~\cite{pystella}.

The scaling parameter, the number of strings per Hubble patch, is defined by $\xi=\frac{l_s t^2}{V a^2}$, where $l_s, V$ denote the total string length and the simulation volume defined in comoving spatial coordinates.
With the scaling parameter at hand, the effective string tension for global and local cases are defined by
\begin{align}
    \label{eq:tension}
    &\mu_{g}=\pi v^2\ln\left(\frac{m_s}{H}\frac{\gamma}{\sqrt{\xi}}\right), \quad
    &\mu_{l}=\pi v^2\ln \left(\frac{m_s}{m_v} \gamma\right)\;,
\end{align}
where $\gamma$ is an additional dimensionless parameter associated with the string configuration during the scaling regime, and it will be taken as a constant in subsequent calculations. The first string tension for the global case is shown in Ref.~\cite{Gorghetto_2021}, and the second corresponds to the modified local string tension with the parameter $\gamma$. Since the denominators in the logarithmic term of the tension are determined by the large-scale cutoff, we adopt the global form when $m_v<H$ and the local form when $m_v>H$, as in Ref.~\cite{2212.13573}. Then we use the energy emission power for a string network in the scaling regime to determine the parameter $\gamma$ in Eq.~\ref{eq:tension}. A more detailed analysis is provided in the {\it Appendix}.

In numerical simulations, computational resources limit the number of grid points, which consequently restricts the frequency range of the resultant GW spectrum. However, assuming a power-law behavior, it is feasible to extend the spectrum to broader frequencies by using the GW emission power $P_g(t)$~\cite{Gorghetto_2021}:
\begin{align}
    \label{eq:rho_gw_with_F}
    \frac{\partial\rho_g}{\partial k}(k,t)&=\int dt^{\prime}\frac{P_g(t^{\prime})}{H(t^{\prime})}\left(\frac{a(t^{\prime})}{a(t)}\right)^3 F\left(\frac{k^{\prime}}{H(t^{\prime})}, \frac{m_s}{H(t^{\prime})}\right)
\end{align}
where $k^{\prime}=ka(t)/a(t^{\prime})$ is the red-shifted momentum. The universal transfer function $F(\frac{k}{H}, \frac{m_s}{H})$, extracted from simulations, captures the power-law structure of the GW spectrum. Assuming $F(x,y)$, with $x=\frac{k}{H}, y=\frac{m_s}{H}$, remains unchanged throughout cosmic history, integrating Eq.\eqref{eq:rho_gw_with_F} over an extended period yields a GW spectrum spanning a broader frequency range. Details of the GW emission power $P_g$ and the function $F(x,y)$ are provided in the {\it Appendix}.

Within the NG approximation, intersections of long, straight strings typically produce loops characterized by kinks and cusps, serving as sources of GW emission. As a loop oscillates, the total GW spectrum is given by the superposition of contributions from all oscillation modes, $\Omega_{\text{gw}}(f)=\sum_j \Omega_{\text{gw}}^j(f)$. For the $j$-th mode, the spectrum behaves as $\Omega_{\rm gw}^{j}\propto j^{-q}$, where $q= 4/3, 5/3, 2$ for GW induced by cusps, kinks and kink-kink collisions respectively. The full GW spectrum is obtained by summing the contributions from all loops in the network, and the detailed procedure is presented in the {\it Appendix}. We note that the BOS and LRS models provide the number density $n$ for NG strings that rely on statistical rules requiring numerous loops.

\noindent{\it \bfseries Numerical results.}
In our simulations, we set $\lambda=0.2, v=2\times10^{17}\textnormal{GeV}$. To investigate the effects of different gauge coupling constants, we select values of $e=0.0005$ (near-global scenario), $e=0.05$, and $e=\sqrt{\lambda/2}\approx0.316$ (equal scalar and gauge field masses). For the case $e=0.0005$, we have $m_v<H$, which leads to the global form of the tension. In the simulation, we introduce dimensionless variables for the field and spacetime coordinates by the rescaling:
\begin{equation}\label{eq:rescale}
    \tilde{\Phi}=\frac{\Phi}{f_*}, \quad \tilde{x}=w_*x
\end{equation}
where $f_*=v$ and $w_*=a_0H_0$, with $a_0=1$. Here $H_0$ is only used to set the reference Hubble scale for the dimensionless simulation variables; equivalently, it corresponds to a reference radiation-dominated background with $T\sim 8\times10^{17}\,\mathrm{GeV}$. With the choice of $a_0$ and $\tilde{\tau}_0=1$, the radiation-dominated background is implemented as $a(\tilde{\tau})=\tilde{\tau}$ in the numerical evolution. The dimensionless comoving grid interval $d\tilde{x}$ is chosen such that the physical interval $a(\tau)dx$ remains smaller than the string widths $r_s$ and $r_v$, defined in Eq.~\eqref{eq:string_width}, throughout the simulation. In addition, to ensure the accuracy of the GW spectrum calculated via the NG method, sufficient loop numbers are necessary, which requires the simulation volume to include an adequate number of Hubble volumes. Thus, we use grid intervals $d\tilde{x}=0.1, d\tilde{t}=0.02$ for GW spectrum calculations and finer intervals $d\tilde{x}=0.05, d\tilde{t}=0.01$ for particle spectrum calculations. The influence of different intervals is discussed further in the {\it Appendix}. We use the total lattice number $N=1024^3$ in the following simulations. 

\begin{figure}[htbp]
    \begin{subfigure}{0.45\textwidth}
    \includegraphics[width=\textwidth]{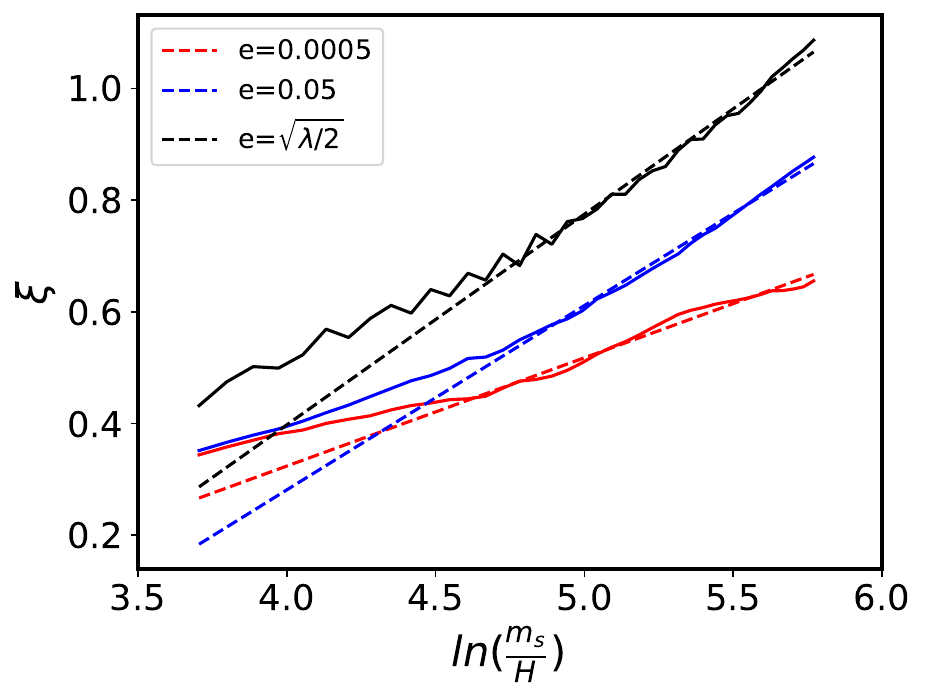}
    \end{subfigure}
    
    \caption{Evolution of the scaling parameter $\xi$ for different gauge couplings. The dashed lines indicate the expected logarithmic behavior $\xi\propto \ln(m_s/H)$.}
    \label{fig:xi_evolution}
\end{figure}

After completing the simulations, we identify the strings using the method described in Ref.~\cite{2311.02011} and the evolution of the scaling parameter $\xi$ is shown in Fig.\ref{fig:xi_evolution}. The dashed lines indicate the expected logarithmic behavior, $\xi \propto \ln(m_s/H)$, which is observed in all cases and is consistent with previous findings for local strings~\cite{Hindmarsh_2017, Hindmarsh_2019, 2212.13573} and global strings~\cite{Gorghetto_2018,2007.04990}.\footnote{This finding is different from Ref.~\cite{PhysRevLett.54.1868} where a constant scalar parameter is observed.} We also find that $\xi$ increases with the gauge coupling $e$, indicating that the gauge coupling affects the density and configuration of the string network. This check is necessary before comparing with the NG prediction, because both the effective string tension and the loop-based reconstruction depend on the scaling properties of the network. 

We then determine the parameter $\gamma$ in Eq.\ref{eq:tension} using field energy emission $P_f$ and GW emission $P_{gw}$ with the method mentioned in the {\it Appendix}. In our analysis, we set $\sqrt{4\pi}\gamma=1.0$ for the near-global scenario and $\sqrt{4\pi}\gamma=0.8, 4.2$ for the local scenarios with $e=0.05, \sqrt{\lambda/2}$, respectively. Additionally, the large fitted value of $\gamma$ for $e=\sqrt{\lambda/2}$ suggests that the effective tension and the network configuration are more strongly modified in the strongly coupled local case.

Having determined the effective string tension, we can compute the GW spectrum using the NG method and compare it with the results obtained from lattice simulations. In the NG framework, cusps, kinks, and kink--kink collisions all contribute to GW emission. However, in field-theoretic simulations the finite string width cannot be neglected, which prevents an exact realization of sharp kinks and cusps. Therefore, we do not distinguish between these localized structures on string loops and instead assume that the GWs are sourced by only one type of feature at a time. With the spectra obtained from both lattice simulations and the NG approximation, we quantify their differences using the function $f_{\text{diff}}$, defined as
\begin{equation}
    \label{eq:diff_lattice_NG}
    f_{\text{diff}}=\sqrt{\frac{1}{N_k}\sum_kf_{\text{diff}}(k)}, \quad f_{\text{diff}}(k)=\frac{(x_l(k)-x_{NG}(k))^2}{x_l(k)^2}
\end{equation}
where $x=\log_{10}{\Omega_{gw}}$ and $x_l, x_{NG}$ denote the results obtained from the lattice simulation and the NG method, respectively. In evaluating $f_{\text{diff}}$, we restrict the summation to the range $10^{-1.3}<k/m_s<1$. This range is chosen to focus on the power-law region of the spectrum, while excluding the infrared part, where the finite simulation volume leads to relatively large statistical uncertainties, and the ultraviolet part, where the spectrum becomes sensitive to the finite string width. This function provides a normalized measure of the difference between the two spectra on a logarithmic scale. Although the fitting range excludes the deep UV region, the strongly coupled local case already has its spectral peak close to $k\sim m_s$, so the intermediate power-law region is still influenced by the microscopic core scale. The resulting values of $f_{\text{diff}}$ are listed in Table~\ref{tab:diff}. For the near-global case with $e = 0.0005$ and the local case with $e = 0.05$, all three source choices give $f_{\text{diff}}<0.1$, indicating good agreement between the lattice results and the NG prediction in the power-law region. The smallest deviation is obtained for kink–kink collisions in the near-global case, while the cusp-induced spectrum gives the best agreement for $e=0.05$. This suggests that the dominant localized structures responsible for GW emission may depend on the gauge coupling, with kink–kink collisions favored in the near-global regime and cusps becoming more important in the local case. In contrast, once $m_v/m_s \gtrsim 1$, the value of $f_{\text{diff}}$ increases significantly for all three types of sources, establishing a \textit{quantitative breakdown criterion} for the NG approximation.

\begin{table}[h]
\centering
\caption{Average differences $f_{\text{diff}}$ between lattice results and NG method results}
\begin{tabular}{|c|c|c|c|}
\hline
 $f_{\text{diff}}$ & $e=0.0005$ & $e=0.05$ & $e=\sqrt{\lambda/2}$ \\ \hline
 cusp & 0.092 & 0.035 & 0.158 \\ \hline
 kink & 0.068 & 0.048 & 0.154 \\ \hline
kink-kink collision & 0.043 & 0.065 & 0.147 \\ \hline
\end{tabular}
\label{tab:diff}
\end{table}

\begin{figure}[!htp]
    \includegraphics[width=0.45\textwidth]{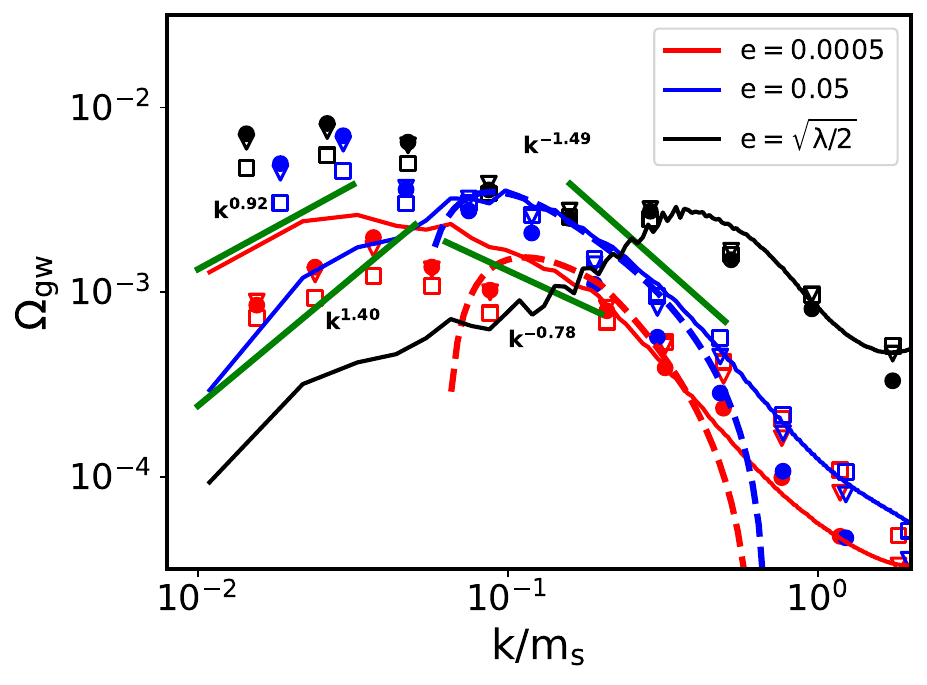}
    \caption{GW spectra for different gauge couplings at $\ln(m_s/H)=5.5$. Solid lines show the spectra obtained from lattice simulations, while dashed lines show the spectra reconstructed using the transfer function $F(x,y)$. Circles, inverted triangles, and squares denote the NG spectra assuming GW emission dominated by kink--kink collisions, kinks, and cusps, respectively.}
    \label{fig:GW_compare}
\end{figure}

We plot the GW spectra obtained from the lattice simulation and from the NG approximation in Fig.~\ref{fig:GW_compare}. The solid lines represent the lattice results, while the circles, inverted triangles, and squares denote the NG spectra assuming GW emission dominated by kink–kink collisions, kinks, and cusps, respectively. To obtain the NG results shown above, we choose an appropriate value of the coefficient $\Gamma$ in Eq.~\ref{eq:GW_spectrum_NG} such that $f_{\text{diff}}$ is minimized for each case. Here $\Gamma$ is treated as an effective normalization parameter. Therefore, the comparison primarily tests whether the NG prescription can reproduce the spectral shape in the power-law region, while the fitted value of $\Gamma$ characterizes the effective GW radiation efficiency. For the NG calculation, we denote the standard value $\Gamma \sim 50$ by $\Gamma_0$, and set $\Gamma=0.04\Gamma_0$ for the near-global case, $\Gamma=0.8\Gamma_0$ for $e=0.05$, and $\Gamma=2\Gamma_0$ for $e=\sqrt{\lambda/2}$. 

With these choices of $\Gamma$, the NG spectra reproduce the lattice results in the power-law region for $e=0.0005$ and $e=0.05$, supporting the validity of the NG approximation in these two cases. More specifically, the best agreement is obtained for kink–kink–collision dominated emission in the near-global case, while the cusp-dominated spectrum gives the smallest deviation for $e=0.05$, consistent with the values of $f_{\text{diff}}$ listed in Table~\ref{tab:diff}. The residual discrepancies outside the power-law region can be attributed to numerical limitations: the infrared part is affected by the finite simulation volume and the limited number of momentum bins, while the ultraviolet part is sensitive to the finite lattice spacing and the finite string width. Since our comparison is restricted to the intermediate power-law region, these edge effects do not affect our assessment of the agreement in that region. In contrast, the $e=\sqrt{\lambda/2}$ case exhibits a sizable deviation from all three NG source prescriptions, indicating that the NG approximation becomes unreliable for local strings with large gauge couplings.

The fitted values of $\Gamma$ also provide an estimate of the effective GW radiation efficiency. Compared with the standard NG expectation, the near-global case requires a much smaller effective $\Gamma$, whereas the local cases require larger values, indicating a lower GW radiation efficiency for near-global strings and a higher efficiency for local strings. This difference is not directly visible from the amplitudes in Fig.~\ref{fig:GW_compare}, because the string tension also varies with the gauge coupling. Although local strings radiate GWs more efficiently, their smaller string tension leads to GW spectra with amplitudes comparable to those of near-global strings. To isolate the effect of the radiation efficiency, we compare global and local strings with the same tension in the {\it Appendix}, where the GW spectrum of the local string is found to be approximately one order of magnitude larger.

As shown in Fig.~\ref{fig:GW_compare}, the GW spectrum shifts toward higher momenta as the gauge coupling increases. The peak of the lattice spectrum moves from $k/m_s \sim 5\times 10^{-2}$ for $e=0.0005$ to $k/m_s \sim 10^{-1}$ for $e=0.05$, and further to $k/m_s=\mathcal{O}(1)$ for the strongly coupled local case with $e=\sqrt{\lambda/2}$. This indicates that the dominant GW emission is no longer generated by structures much larger than the string core, but is instead closely tied to small-scale structures of the field-theoretic string network. This observation explains why a more localized string is not necessarily better described by the NG approximation. The NG approximation treats a string as an infinitely thin object and therefore neglects the internal field structure of the string core. It is reliable only when the relevant GW-emitting structures are much larger than the microscopic string width. In the strongly coupled local case, increasing the gauge coupling makes the string energy more localized, but it also drives the dominant GW-emitting modes toward the core scale. As a result, finite-width effects and the additional gauge-field degrees of freedom can no longer be neglected. By contrast, in the near-global case, the dominant GW-emitting modes have wavelengths much larger than the string width. Therefore, despite the presence of long-range field gradients, the microscopic core structure is less directly probed by GW emission, allowing the NG approximation to reproduce the lattice spectrum in the power-law region. A more detailed investigation of the discrepancy in the strongly coupled local case will be carried out in future work.

With the GW spectrum calculated from the lattice simulations, the universal transfer function $F(x,y)$ can be derived, and a new spectrum can be computed using Eq.~\eqref{eq:rho_gw_with_F}. The resulting transfer function $F(x,y)$ is shown in the {\it Appendix}, which can be used for future lattice-calibrated extrapolations of string-induced GW spectra. Subsequent analyses focus exclusively on the cases $e=0.0005$ and $e=0.05$, where the spectra exhibit power-law behavior. To verify this approach, we restrict the integration to our simulation period, presenting results at $\ln(\frac{m_s}{H})=5.5$ in Fig.\ref{fig:GW_compare}, which confirms that the GW spectrum derived from $F(x,y)$ aligns closely with our previous results. 

\begin{figure}[htbp]
    \includegraphics[width=0.45\textwidth] {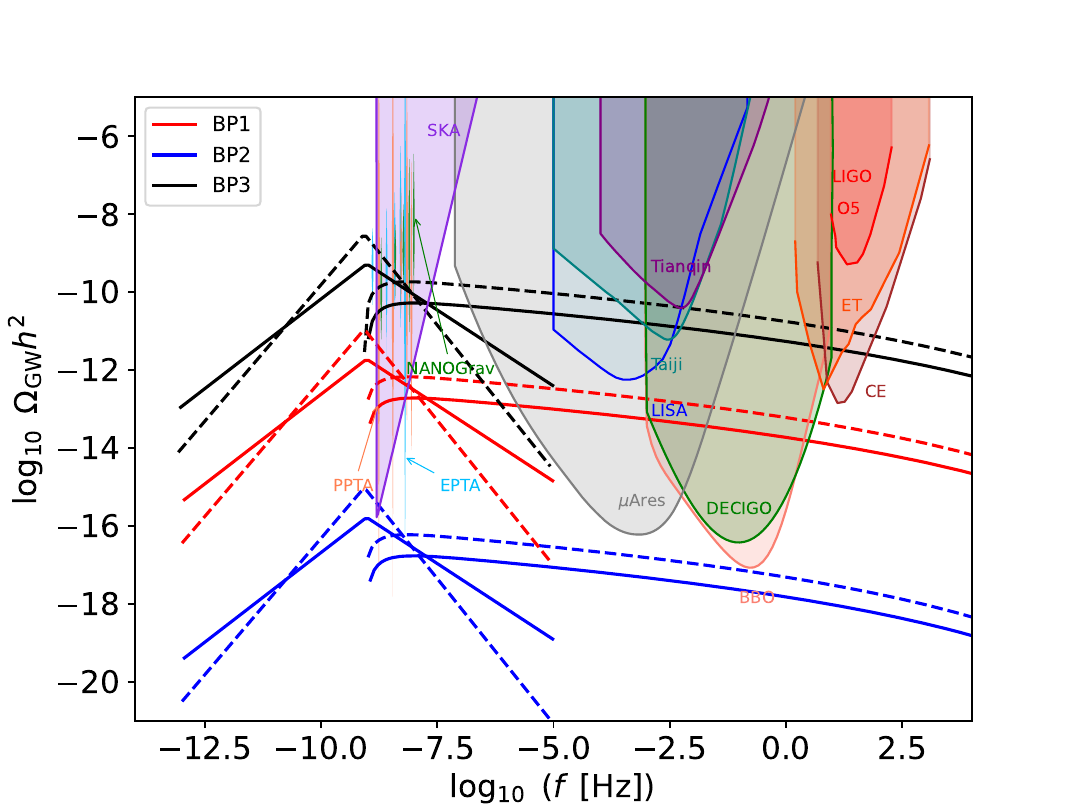}

    \caption{Detectability of the GW spectra for three benchmark cases. We adopt $f_a = 5\times10^{14} ～\text{GeV}$, $5\times10^{13} ～\text{GeV}$, and $2\times10^{15} ～\text{GeV}$ for BP1, BP2, and BP3, respectively. The solid and dashed curves correspond to $e = 0.0005$ and $e = 0.05$, respectively. The sensitivity curves of LISA~\cite{gw_LISA_1, gw_LISA_2}, TianQin \cite{gw_TianQin_1, gw_TianQin_2}, Taiji \cite{gw_TaiJi_1, gw_TaiJi_2}, $\mu$Ares \cite{gw_muAres_1}, DECIGO \cite{gw_DECIGO_1,gw_DECIGO_2,gw_DECIGO_3,gw_DECIGO_4}, BBO \cite{gw_BBO_1,gw_BBO_2,gw_BBO_3}, SKA \cite{gw_SKA_1}, ET \cite{gw_ET_1, gw_ET_2}, CE \cite{gw_CE_1}, LIGO \cite{gw_LIGO_1,gw_LIGO_2,gw_LIGO_3}, NANOGrav \cite{gw_NANOGrav_1}, PPTA \cite{gw_PPTA_1} and EPTA \cite{gw_EPTA_1} are presented.}
    \label{fig:GW_experiment}

\end{figure}

To extend the GW spectrum beyond the limited dynamical range of the lattice simulations, we use two complementary procedures. First, using the peak frequency, the spectral slopes on both sides of the peak, and the total GW power directly obtained from the lattice simulations, we estimate the present-day spectra associated with the simulated frequency range. Second, we use the transfer function $F(x,y)$ extracted from the simulations and integrate Eq.\eqref{eq:rho_gw_with_F} up to the time when the string network collapses, thereby obtaining spectra with a broader frequency coverage. In the benchmark calculation, the collapse time is chosen to be $H=5\times 10^{-23}\text{GeV}$, corresponding to $T\sim 6~\text{MeV}$. The spectra obtained from both procedures are redshifted to the present day and shown in Fig.\ref{fig:GW_experiment}, together with the sensitivity curves of current and future GW experiments. It can be seen that the spectrum corresponding to $f_a=2\times10^{15}\text{GeV}$ could be probed by current PTA experiments and future space-based GW detectors, including LISA, TianQin, and Taiji. The amplitude of the spectra appears smaller than that reported in \cite{2212.13573}, where the calculation is based on the NG approximation. We note that our analysis assumes a radiation-dominated Universe, and the integration of Eq.\ref{eq:rho_gw_with_F} terminates before the radiation–matter equality. In contrast, the NG string calculation in \cite{2212.13573} neglects the influence of the cosmological background and assumes continuous emission from string loops, with some long-lived loops radiating until $z \sim 1$. As a result, their spectra extend over a longer emission period and exhibit a higher GW amplitude.

\begin{figure}[htbp]
    \includegraphics[width=0.45\textwidth] {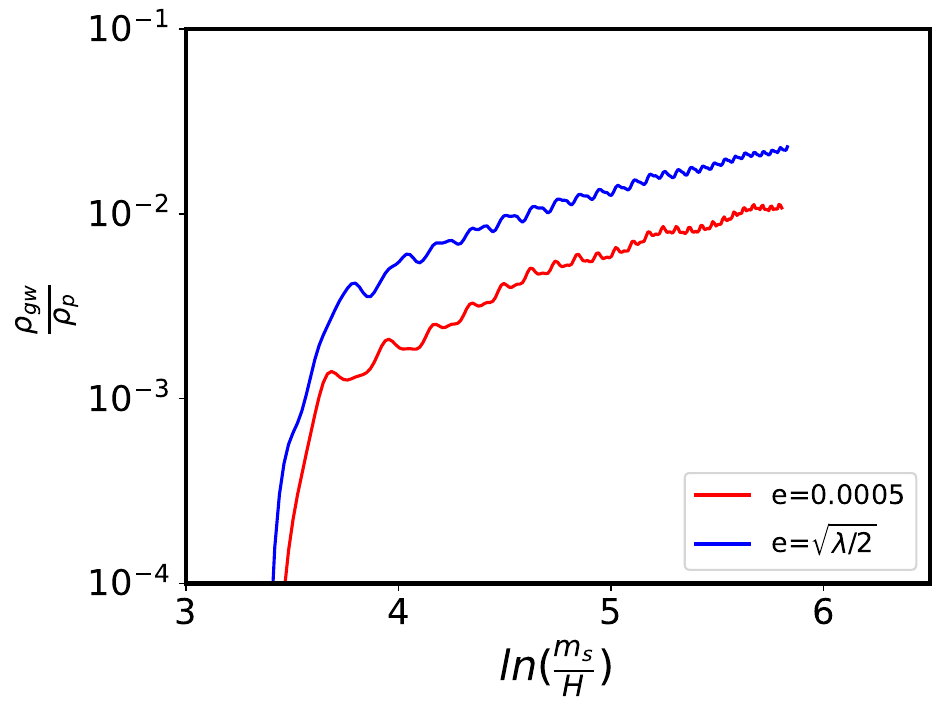}

    \caption{Ratio of GW energy density $\rho_{gw}$ to particle energy density $\rho_p$. Particle emission dominates by more than two orders of magnitude, showing that GW radiation is subdominant to particle emission.}
    \label{fig:ratio_GW_particle}

\end{figure}

\begin{figure}[htbp]
    \begin{subfigure}{0.45\textwidth}
    \includegraphics[width=\textwidth] {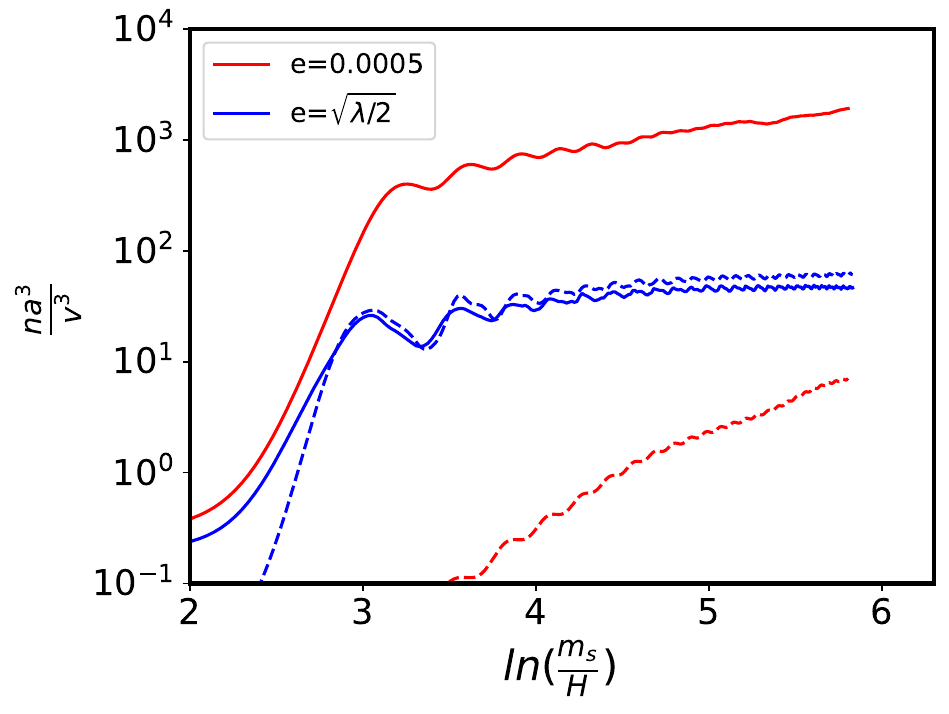}
    \end{subfigure}
    \begin{subfigure}{0.45\textwidth}
    \includegraphics[width=\textwidth]{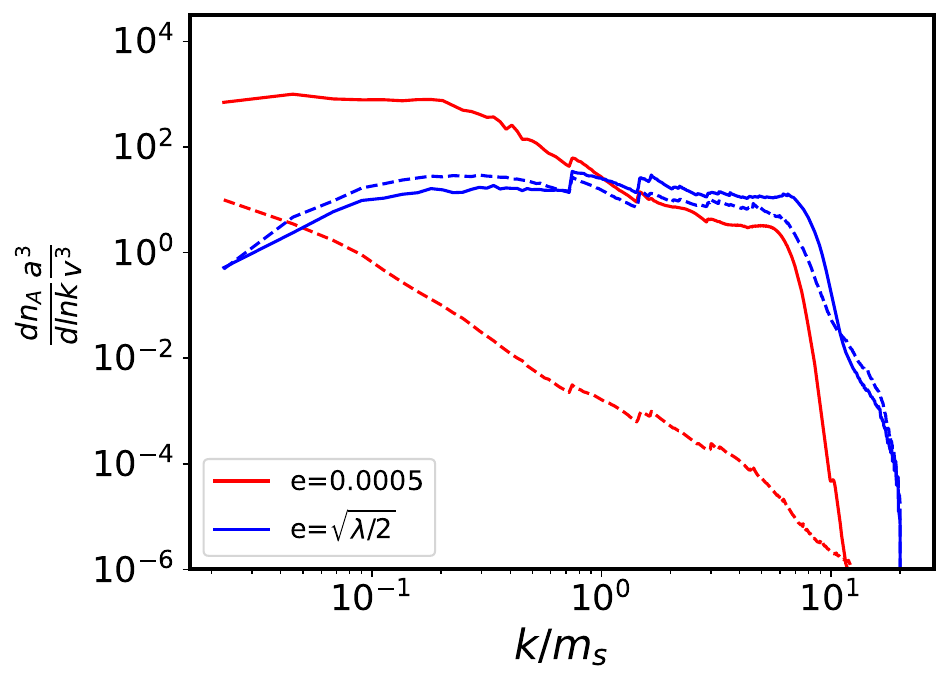}
    \end{subfigure}
    
    \caption{Left: Evolution of particle number density. Right: Particle spectra. Solid and dashed lines correspond to longitudinal and transverse polarizations, respectively.}
    \label{fig:particle_number}
\end{figure}

Finally, we turn to the particle-emission channel, whose detailed calculation is given in the {\it Appendix}. The ratio between the GW energy density $\rho_{\rm gw}$ and the particle energy density $\rho_p$ is shown in Fig.~\ref{fig:ratio_GW_particle}. Due to computational resource constraints associated with smaller time intervals requiring more simulation steps, we present results exclusively for the cases $e = 0.0005$ (near-global) and $e = \sqrt{\lambda/2}$ (local), highlighting differences between global and local scenarios. The GW emission is approximately $O(10^{-3}) - O(10^{-2})$ of the particle emission energy, aligning closely with previous results reported in Ref.~\cite{2412.04218v1}. The ratio is slightly higher for the local case despite its smaller string tension. The increase in this ratio indicates that GW emission continues after the growth of the particle number density slows down, consistent with Fig.~\ref{fig:particle_number}.

We further decompose the particle spectrum into longitudinal and transverse polarizations, as described in the {\it Appendix}. Fig.\ref{fig:particle_number} displays the evolution of the particle number density and the particle spectra, both rescaled by $a^3/v^3$ to form dimensionless quantities and to account for the dilution due to cosmic expansion.

In the near-global case, the small gauge-boson mass allows light field modes to be efficiently excited by the evolving string network, leading to a larger particle number density. In the local case, the larger gauge-boson mass suppresses the excitation of low-momentum modes, reducing the total particle number density. Additionally, the gauge coupling constant governs the energy transfer efficiency between scalar and gauge fields, and the transverse component originates exclusively from the gauge field. Thus, the number density of the longitudinal polarization substantially exceeds that of the transverse polarization in the near-global scenario, while the difference between them is marginal in the local scenario.

\noindent{\it \bfseries Conclusions.} We quantitatively determine the regime of validity of the NG approximation by comparing GW spectra from zero-temperature Abelian-Higgs lattice simulations with those predicted by the NG description. The NG approximation reproduces the lattice spectra in the power-law region for the near-global case and for the local case with moderate gauge coupling, but breaks down for strongly coupled local strings with $m_v/m_s\sim 1$. This behavior shows that the validity of the NG approximation for GW predictions is governed not only by the localization of the string energy, but also by the characteristic scale of the dominant GW-emitting modes. In the near-global case, the relevant modes have wavelengths much larger than the string width, whereas in the strongly coupled local case they probe the string-core scale, where finite-width effects and gauge-field degrees of freedom cannot be neglected. Our results help interpret stochastic GW backgrounds predicted by NG-based models and provide a lattice-calibrated connection between particle model parameters and GW detectability. We find that the GW spectra for the symmetry breaking scale $f_a=2\times 10^{15}~\mathrm{GeV}$ can be probed by PTAs, LISA, TianQin, and Taiji. For the same string tension, the GW spectrum of a local string is approximately one order of magnitude larger than that of a global string. Furthermore, our results confirm that particle radiation dominates over GW emission as the main energy-loss channel of the string network, establishing a useful numerical benchmark for future observational and numerical studies of particle emission from cosmic strings.

\noindent{\it \bfseries Acknowledgments.}
This work is supported by the National Key Research and Development Program of China under Grant No.2021YFC2203004, and by the National Natural Science Foundation of China (NSFC) under Grants Nos.12322505, 12547101. We also acknowledge Chongqing Talents: Exceptional Young Talents Project No. cstc2024ycjhbgzxm0020 and Chongqing Natural Science Foundation under Grant No. CSTB2024NSCQJQX0022. J.S. is supported by the Peking University under startup Grant No.7101302974, the NSFC under Grants No.12025507, No.12450006.

\bibliographystyle{apsrev4-1}
\bibliography{references}  

\clearpage

\onecolumngrid
\begin{center}
  \textbf{\large Appendix}\\[.2cm]
\end{center}
This Appendix provides the initial conditions of the fields, the calculation of the effective string tension and the GW emission power for NG strings, technical details of the particle-emission calculation, the universal transfer function analysis, a comparison of global and local strings with equal tension, and checks on the effect of lattice resolution.

\noindent{\it \bfseries Initial conditions for the scalar field.} A quantum spectrum is used to characterize the distribution of the complex scalar field and its conjugate momenta in momentum space. The initial scalar field values satisfy:
\begin{align}
    \label{eq:init_scalar}
    \langle\phi(\mathbf{k})\phi(\mathbf{k}^{\prime})\rangle &= (2\pi)^3P_{\phi}(|\mathbf{k}|)\delta^3(\mathbf{k}-\mathbf{k}^{\prime}), \notag \\
    \langle\pi(\mathbf{k})\pi(\mathbf{k}^{\prime})\rangle &= (2\pi)^3P_{\pi}(|\mathbf{k}|)\delta^3(\mathbf{k}-\mathbf{k}^{\prime})
\end{align}
where $\phi(k)$ denotes either the real or imaginary component of the scalar field $\Phi$ in momentum space, and $\pi(k)$ represents its conjugate momentum. The power spectra $P(k)$ are defined as follows:
\begin{align}
    \label{eq:init_scalar_spectrum}
    P_{\phi}(k) &= \frac{1}{\omega_k}, \quad
    P_{\pi}(k) = \omega_k,
\end{align}
with $\omega_k=\sqrt{k^2+m_{s}^2}$, where $m_{s}$ denotes the mass of scalar field.

\noindent{\it \bfseries Calculation of emission power and string tension.} The emission power for a global string network in the scaling regime is defined as the difference in power between a 'free' network, where strings simply stretch and dilute due to cosmic expansion, and the 'real' network, which has reached the scaling regime. The emission power given by \cite{Gorghetto_2021} is 
\begin{align}\label{eq:power_loss_global}
    P(t_0)&=[\dot{\rho}^{\text{free}}(t)-\dot{\rho}(t)]_{t=t_0} =\rho(t_0)\left[2H-\frac{\dot{\xi}}{\xi}-\frac{\mu_0}{\mu}\left(H+\frac{\dot{\gamma}}{\gamma}-\frac{\dot{\xi}}{2\xi}\right)\right]\notag \\
    &\overset{\ln(m_r/H)\gg 1}{\longrightarrow} 8H^3(t)\xi(t)\mu(t),
\end{align}
Here, the global string tension defined in Eq.\eqref{eq:tension} has been used, and the scaling parameter is assumed to follow $\xi\propto\ln(\frac{m_s}{H})$. This analysis also remains valid for local string networks. Assuming $\xi$ is still proportional to $\ln(\frac{m_s}{H})$, the only difference is the form of string tension. Since there is no explicit $H$-dependence in this definition, the equation for the energy emission power simplifies to:
\begin{align}\label{eq:power_loss_local}
    P(t_0)&=[\dot{\rho}^{\text{free}}(t)-\dot{\rho}(t)]_{t=t_0} =\rho(t_0)\left(2H-\frac{\dot{\xi}}{\xi}\right)\notag \\
    &\overset{\ln(m_r/H)\gg 1}{\longrightarrow} 8H^3(t)\xi(t)\mu.
\end{align}
Hence, under the long-time limit, the expression resembles the global string case. This allows us to extract the string tension $\mu$ directly from simulations by evaluating the energy emission power and the scaling parameter $\xi$.

To determine the effective string tension (\ref{eq:tension}), we first calculate the emission power of the string network. The total energy emission power $P_{tot}=P_f+P_{gw}$, where $P_f$ is field energy emission from string network and GW emission $P_{gw}$ is derived from the lattice GW spectrum. Then we adjust the parameter $\gamma$ in Eqs.\eqref{eq:tension} until the theoretical prediction $P=8H^3\xi\mu$ aligns with $P_{tot}$, ensuring that the ratio $P_{tot}/P$ stabilizes at unity. The field emission power of the cosmic string can be calculate by field energy $\rho_f$ in the lattice
\begin{align}\label{eq:power}
    P_{i}=\int dk \frac{1}{a^{z(k)}}\frac{\partial}{\partial t} \left(a^{z(k)}\frac{\partial\rho_i}{\partial k}\right),
\end{align}
where $i$ represents each field, and $z(k)=3+(k/m_i)^2/((k/m_i)^2+1)$ is the momentum-dependent redshift factor. For massless fields, such as the scalar phase in the near-global case, $z(k)=4$.

\begin{figure}[htbp]
    \begin{subfigure}{0.4\textwidth}
    \includegraphics[width=\textwidth]{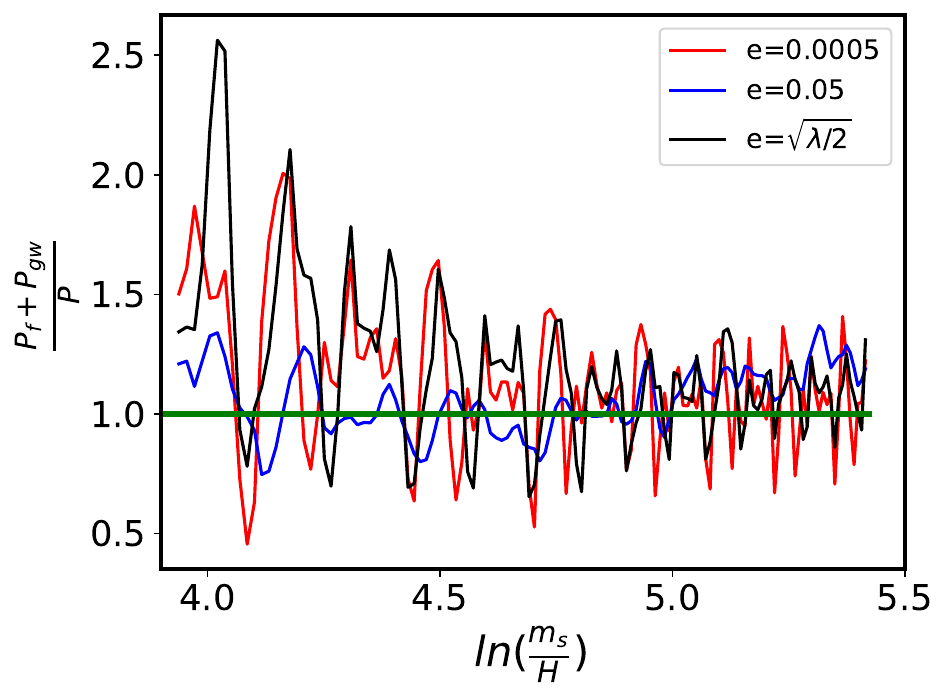}
    \end{subfigure}
    \begin{subfigure}{0.4\textwidth}
    \includegraphics[width=\textwidth]{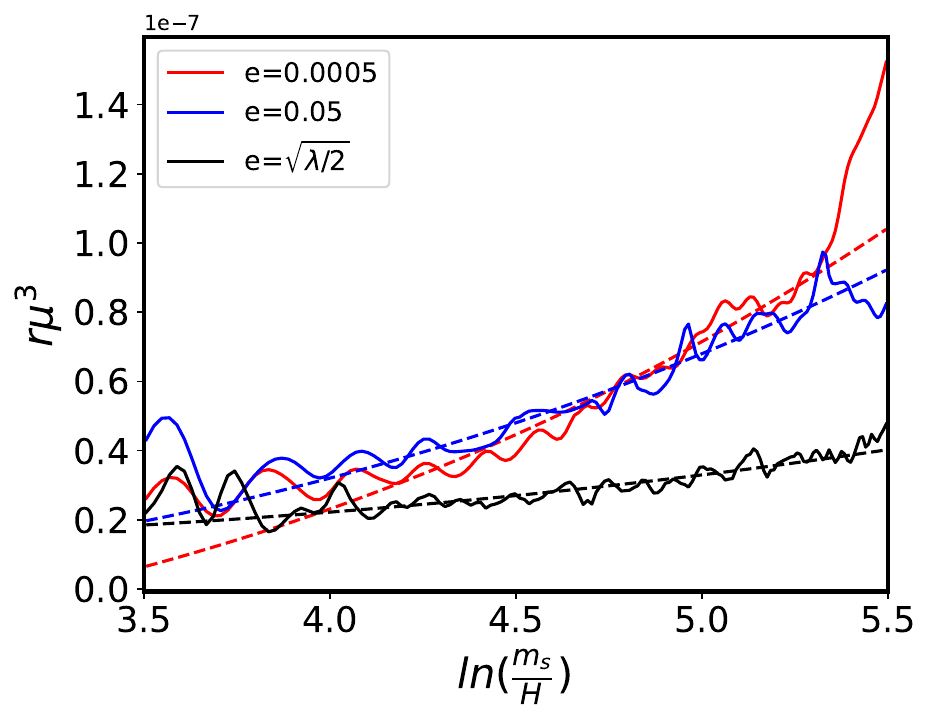}
    \end{subfigure}

    \caption{Left: the ratio $P_{tot}/P$. Right: the parameter $r\mu^3$, with fitted results indicated by dashed lines.}
    \label{fig:Gamma}
\end{figure}

With $\sqrt{4\pi}\gamma=1.0, 0.8 \text{ and } 4.2$ for $e=0.0005, 0.05 \text{ and } \sqrt{\lambda/2}$, the ratios are depicted in the left panel of Fig.\ref{fig:Gamma}. Because the power of field oscillates strongly in simulation, we take the average value in $\Delta \ln(\frac{m_s}{H})=0.4$ to get this figure. 

To extend the GW spectrum calculated via lattice method using the equation \ref{eq:rho_gw_with_F}, we first need to get a analytical expression of GW power $P_g(t)$ from the numerical result. We use the method mention in Ref.~\cite{Gorghetto_2021}. From dimensional analysis and the NG approximation, field emission power should scale with the square of the VEV of the field, $v^2$, and GW emission power should scale with $G\mu^2$
\begin{equation}\label{eq:energy_emission_power}
    \frac{dE_{gw}}{dt}=r_g G\mu^2, \quad \frac{dE_f}{dt}=r_f v^2.
\end{equation}
If $P_{gw}\ll P_f$, the total emission power approximates $P\approx P_f$. Therefore, in the long-time limit, the GW emission power becomes:
\begin{equation}\label{eq:ratio_P}
    P_g(t)=r\frac{G\mu^2}{v^2}P(t)=\frac{8Gr\mu^3H^3(t)\xi(t)}{v^2},
\end{equation}
where $r$ is a functional dependent on the average shape of the string network. Now lattice simulation gives $P_g(t)$ and string tension $\mu$ has been known, so this equation can be used to calculate the ratio $r$ and then we can use this equation as analytical expression of $P_g(t)$.

However, from Eq.\eqref{eq:ratio_P}, the key combination is $r\mu^3$ rather than $r$ alone. Since plotting  individually would yield distinctly different evolutions for global and local scenarios due to their differing $\mu$ forms, we focus instead on the quantity $r\mu^3$, whose evolution is shown in the right panel of Fig.\ref{fig:Gamma}, and which is used in subsequent analyses. From the result in \cite{Gorghetto_2021}, where $r$ is constant and $\mu\propto \ln(\frac{m_s}{H})$ for global strings, they get $r\mu^3\propto (\ln(\frac{m_s}{H}))^3$. We extend this result to the local string scenario, using the form $r\mu^3\propto (\ln(\frac{m_s}{H}))^3$ to fit our numerical results, as indicated by dashed lines in the figure.

\noindent{\it \bfseries Calculation of GW spectrum for NG strings} Within the NG approximation, NG strings intersect and reconnect, forming string loops that oscillate with a period $T=L/2$, where $L$ denotes the length of the loop. Intersections of long, straight strings typically produce loops characterized by kinks and cusps, serving as sources of GW emission. For the j-th oscillation mode, the frequency is $f_j=2j/L$, and the corresponding GW spectrum scales as $\Omega^j_{gw}\propto j^{-q}$, where $q=4/3, 5/3, 2$ for GW emission induced by cusps, kinks and kink-kink collisions, respectively. The GW emission power of a single loop can be approximated by $\dot{E}_{gw}=\Gamma G\mu^2$ \cite{texbook_cosmic_string}, where $\mu$ is the string tension can be  obtained through Eq.~\ref{eq:tension}, and previous numerical simulations suggest $\Gamma\sim 50$. Considering a string network characterized by the loop number density $n(t,l)$, which indicates the loop density with length $l$ at time $t$, the GW spectrum for the j-th mode is given by~\cite{2002.01079}:
\begin{align}\label{eq:GW_spectrum_NG}
    \Omega_{\text{gw}}^j(f_0)=&\frac{j^{-q}}{\epsilon}\frac{8\pi}{3}\left(\frac{G\mu}{H_0}\right)^2\frac{2j}{f_0a_0}\Gamma \notag \\
    &\times\int_{a_{\text{min}}}^{a_{\text{max}}}\left(\frac{a}{a_0}\right)^4\frac{da}{H(a)}n\left(t(a),l_j(a)\right),
\end{align}
where $a_0, H_0$ are both defined at present and $\Omega_{gw}^j(f_0)$ represents the GW spectrum for the frequency $f_0$ at present, implying the corresponding frequency at time $t$ is $f(t)=a_0f_0/a(t)$. Additionally, $l_j(a)=2ja/(f_0a_0)$ denotes the loop length, and $\epsilon=\sum_j j^{-q}$ is the normalization factors. The total GW spectrum is obtained by summing contributions from all modes, $\Omega_{\text{gw}}(f)=\sum_j \Omega_{\text{gw}}^j(f)$.

Following lattice simulations, the GW spectrum from string loops can be computed using Eq.~\eqref{eq:GW_spectrum_NG} by statistically analyzing loop number densities at each simulation step. Therefore, upon completing the field evolution, it is necessary to identify loops within the string network. In our implementation, we identify strings using the same method described in \cite{2311.02011}, locating each string by detecting connected regions. We then iterate over each point on a string and calculate the distance between adjacent points. A string loop is characterized by its ends being connected, such that the distance between each pair of adjacent points remains small. Due to periodic boundary conditions, an even number of transitions across the boundary in any direction should not be interpreted as a real discontinuity. Using this method, we are able to reliably identify individual loops and compute their lengths.

For an individual loop of length $l_0$, the number density distribution is expressed as $n(t,l)=\frac{1}{V}\delta(l-l_0)$. Given that the delta function $\delta(l-l_0)$ has the dimension of $l_0^{-1}$, one cannot simply take $\delta(l-l_0)=1$. Instead, for a loop of length $l_0$, the number of loops within the interval $l_0\sim l_0+dl$ is one, implying the number density for a single loop should be defined as $n(t,l)=\frac{1}{Vdl}$ when $l=l_0$, and $n(t,l)=0$ otherwise, where $dl=adx$ is measured on the comoving lattice. The overall GW spectrum obtained using the NG approximation is thus the sum of contributions from each individual loop.

Each string loop emits GWs at specific frequencies that differ slightly over time. However, arbitrarily fine frequency spacing is not physically meaningful in numerical simulations. To maintain the shape of the spectrum in log-log figure, the width of bins should be equal. We select an appropriate momentum interval, setting the frequency bin width to $f_{min}/2$, where $f_{min}$ is the lowest frequency generated by NG loops. The summed spectrum within each frequency bin yields the final GW spectrum.

\noindent{\it \bfseries The local and global string with the same tension.} To investigate whether the comparable GW spectra generated by global and local strings arise from differing tensions rather than equivalent GW emission efficiencies, we calculate the GW spectrum for a modified local case. Specifically, we set $\lambda=6, e=0.01$, resulting in a gauge field mass $m_v^{\prime}>H$ throughout the evolution, thereby characterizing it as a local string scenario. Supposing the parameter $\gamma$ is the same as near-global case, the corresponding string tension $\mu=\pi v^2\ln{(\frac{m_s^{\prime}}{m_v^{\prime}})}$ matches the tension of the near-global case at about $\ln{(\frac{m_s}{H})}=5.1$. In this simulation, the lattice spacing $d\tilde{x}$ is small due to the large new scalar field mass $m_s^{\prime}$ but it still remains super-horizon. 

\begin{figure}[htbp]
    \includegraphics[width=0.45\textwidth]{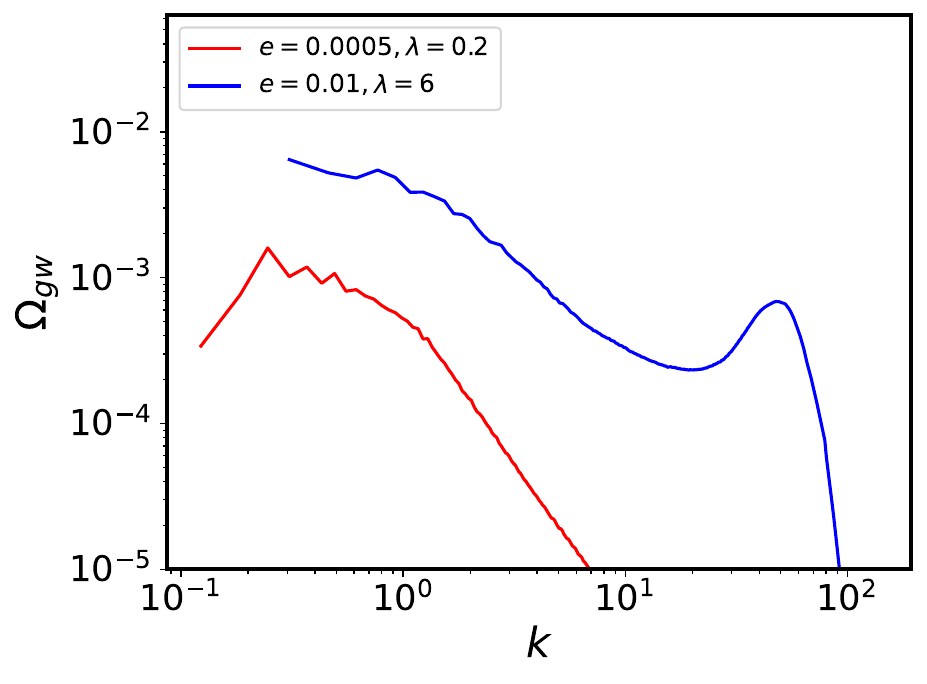}
    \caption{The GW spectrum for the new local case ($e=0.01, \lambda=6$) and the near-global case ($e=0.0005, \lambda=0.2$) at the same time.}
    \label{fig:GW_tension}
\end{figure}

Figure \ref{fig:GW_tension} compares the GW spectra of this modified local scenario and the near-global scenario at the same evolution time. Despite the slightly lower tension in the modified local case, its GW spectrum is noticeably higher, indicating that local strings emit GW more efficiently than global strings. The shift observed along the horizontal axis is attributed to the smaller lattice spacing $d\tilde{x}$ required for the modified local scenario due to its larger scalar field mass.

\noindent{\it \bfseries GW spectrum universal transfer function } With the GW spectrum $\frac{d\rho_{gw}}{dk}$, the universal transfer function in Eq.\ref{eq:rho_gw_with_F} can be calculated by 
\begin{equation}
    F\Big(\frac{k}{H}, \frac{m_s}{H}\Big)=\frac{H}{P_g}\frac{1}{a^3(t)}\frac{\partial}{\partial t}\Big( a^3(t) \frac{\partial\rho_{gw}}{\partial k}(k,t)\Big)
\end{equation}
Following the analysis of Ref.~\cite{Gorghetto_2021}, the function $F(x,y)$ satisfies $\int dxF(x,y)=1$, which can be used to determine the coefficient $\frac{H}{P_g}$. We use the power-law shape spectrum, it can be approximated by 
\begin{equation}
    \label{eq:F}
    F(x,y)=
        \begin{cases} 
        \frac{(q-1)x_0^{q-1}}{x^q}, & x \in [x_0,y], \\
        0, & x \notin [x_0,y]
        \end{cases}
\end{equation}
where $x_0$ and $y$ represent the IR peak and UV cutoff of the GW spectrum, respectively.

\begin{figure}[htbp]
    \includegraphics[width=0.45\textwidth]{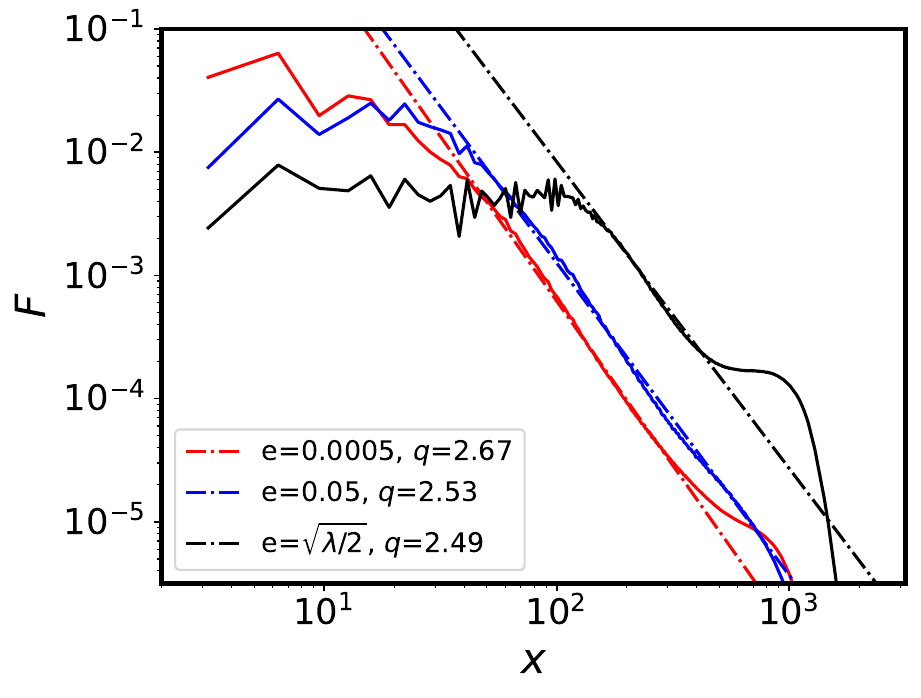}
    \caption{The universal transfer function $F(x,y)$ of GW spectrum. The dashed line is the fitted line.}
    \label{fig:F}
\end{figure}

The results for $F(x,y)$ are depicted in Fig.\ref{fig:F}. For the cases $e=0.0005$ and $e=0.05$, $F(x,y)$ exhibits a clear power-law behavior over the relevant momentum range, supporting the use of the power-law parametrization in Eq.\eqref{eq:F} for the subsequent extrapolation. However, for the case $e=\sqrt{\lambda/2}$, although $F(x,y)$ shows an approximate power-law behavior at high momenta, its overall shape significantly deviates from a single power law, violating the underlying assumption. Thus, we focus only on the cases $e=0.0005$ and $e=0.05$ for the above calculation. In calculating $F(x,y)$, we utilize the mean emission power spectrum from the simulation, rather than instantaneous values.

\noindent{\it \bfseries Calculation of particle emission spectrum}
The energy density radiated by the string network is given by \cite{2212.13573}:
\begin{align}
    \label{eq:energy_field_1}
    \rho_A&=\frac{|\Phi|^2}{v^2}\frac{1}{a^2}\Big[|D_0\Phi|^2+|D_i\Phi|^2+\frac{1}{2a^2}(E_i^2+B_i^2)\Big] \notag \\
    &\approx\frac{|\Phi|^2}{v^2}\left[\frac{2}{a^2}\left(\frac{\textnormal{Im}(\Phi^{\dagger}\Pi)}{|\Phi|}\right)^2+\frac{1}{a^4}(E_i)^2\right] \notag \\
    &=\frac{1}{a^4}\left[\frac{2(\partial_i E_i)^2}{a^2m_v^2}+(\frac{|\Phi|}{v} E_i)^2\right],
\end{align}
where the excitation of the radial component of the scalar field is neglected. We assume, for simplicity, the equality of gradient and kinetic energies for the scalar field, as well as the equality of electric and magnetic energies for the gauge field. Additionally, the Gauss constraint $\partial_iE_i=ea^2\textnormal{Im}(\Phi^{\dagger}\Phi^{\prime})$ is utilized, and an additional factor $\frac{|\Phi|^2}{v^2}$ is introduced to exclude the influence of the string. Then we decompose the vector field $|\Phi|E_i$ to determine the particle spectrum for each polarization:
\begin{equation}
    n_A^{L,T}=\int dk\frac{n_A^{L,T}(k)}{dk}=\int dk\frac{1}{E_A(k)}\frac{\rho_A^{L,T}(k)}{dk},
    \end{equation}
where $E_A(k)=\sqrt{(k/a)^2+m_A^2}$. 

\begin{figure}[htbp]
    \centering
    \begin{subfigure}{0.45\textwidth}
    \includegraphics[width=\textwidth] {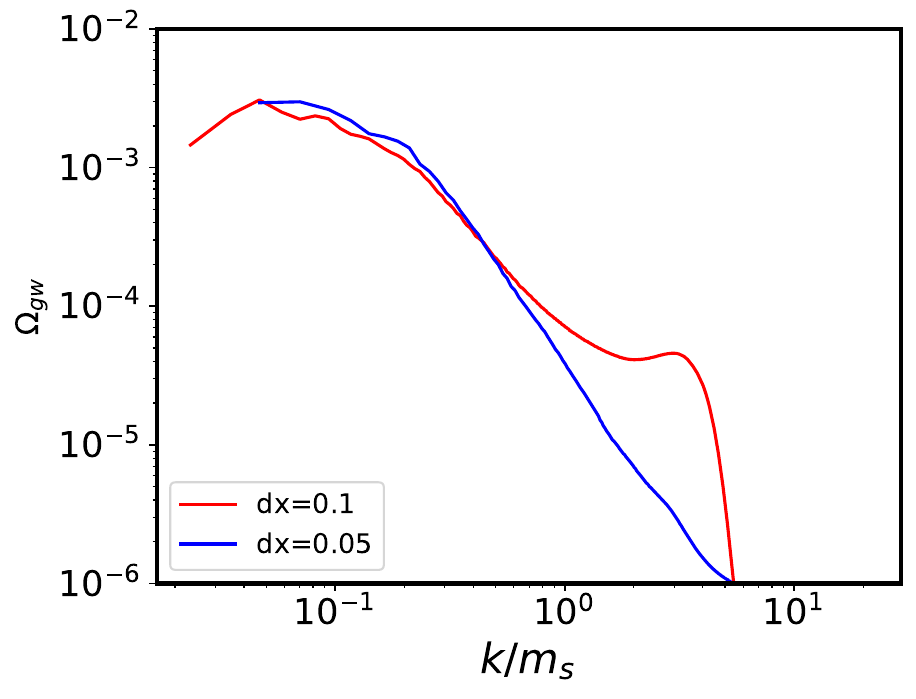}
    \end{subfigure}
    \begin{subfigure}{0.45\textwidth}
    \includegraphics[width=\textwidth]{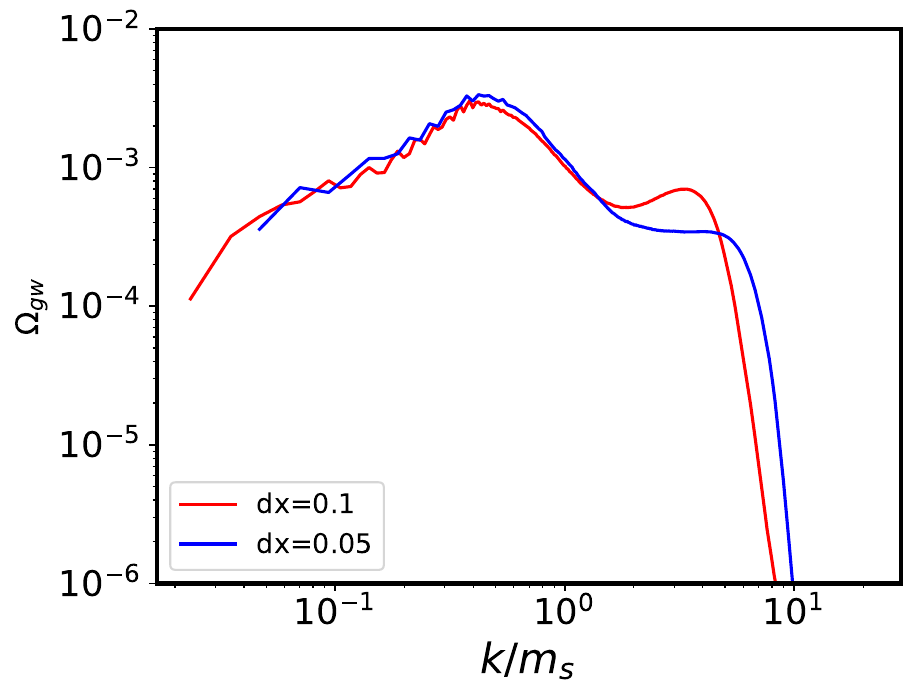}
    \end{subfigure}
    \caption{GW spectra calculated with different $d\tilde{x}$ for $e=0.0005$(left) and $e=\sqrt{\lambda/2}$ (right).}
\label{fig:GW_diff_dx}
\end{figure}

\begin{figure}[htbp]
    \begin{subfigure}{0.45\textwidth}
    \includegraphics[width=\textwidth] {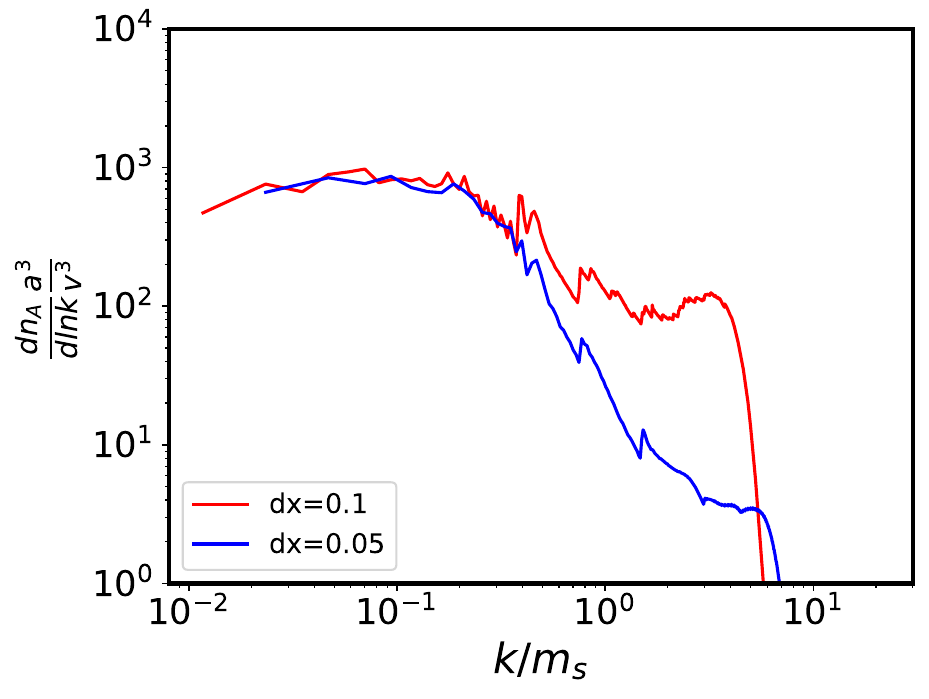}
    \end{subfigure}
    \begin{subfigure}{0.45\textwidth}
    \includegraphics[width=\textwidth]{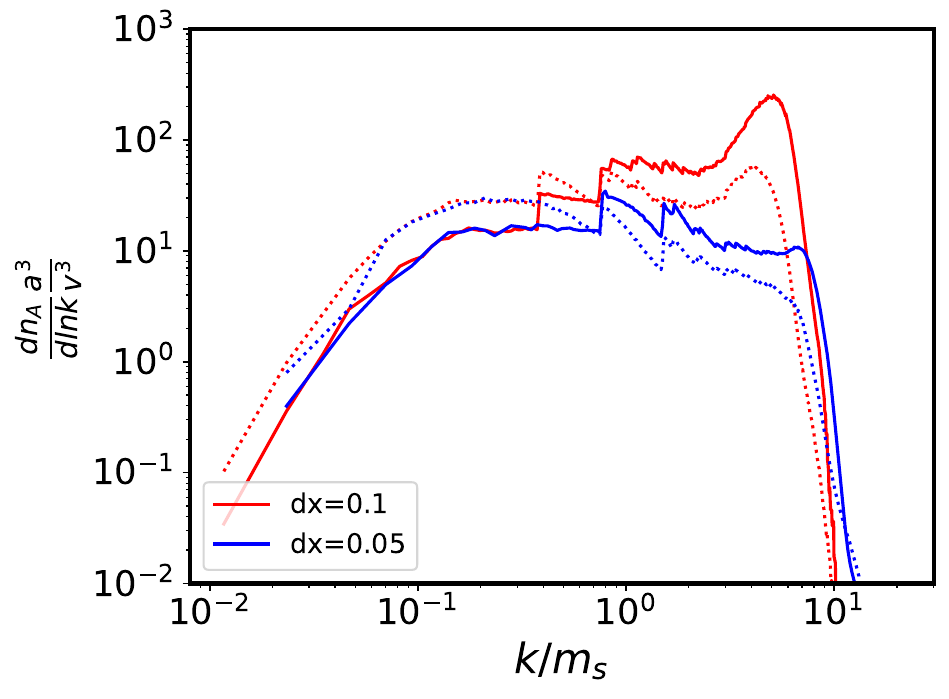}
    \end{subfigure}
    \caption{Particle spectra calculated with different $d\tilde{x}$ for $e=0.0005$(left) and $e=\sqrt{\lambda/2}$ (right). Solid and dashed lines denote longitudinal and transverse polarizations, respectively.}
\label{fig:n_A_diff_dx}
\end{figure}

\noindent{\it \bfseries The influence of $\bm{d\tilde{x}}$} To examine the influence of the spatial resolution $d\tilde{x}$, we perform simulations with the same number of lattice points but two different lattice spacings, $d\tilde{x}=0.1$ and $d\tilde{x}=0.05$. The GW spectra computed from these simulations are shown in Fig.\ref{fig:GW_diff_dx} for the cases $e=0.0005$ and $e=\sqrt{\lambda/2}$, and the corresponding particle spectra are presented in Fig.\ref{fig:n_A_diff_dx}. This comparison provides a resolution check for the spectral results.

Reducing $d\tilde{x}$ suppresses the high-momentum part of both the GW and particle spectra, while leaving the low-momentum behavior largely unchanged. This indicates that the finer lattice resolves the small-scale radiation more accurately, without qualitatively changing the large-scale dynamics of the string network. The stability of the low- and intermediate-momentum spectra under this change of $d\tilde{x}$ supports the robustness of the power-law behavior used in our GW analysis. For the GW spectrum, our main focus is the overall spectral shape and the power-law behavior in the intermediate momentum range. The high-momentum corrections induced by changing $d\tilde{x}$ therefore do not affect our main conclusions. Moreover, the comparison with the NG approximation requires a sufficiently large simulation volume to produce enough string loops for a reliable loop-number distribution. We therefore use $d\tilde{x}=0.1$ for the GW analysis.

By contrast, the particle spectrum is more sensitive to the high-momentum region, because the particle energy density and spectral distribution receive sizable contributions from small-scale radiation. Since decreasing $d\tilde{x}$ strongly affects the high-momentum tail but does not substantially modify the low-momentum behavior, we use the higher-resolution simulations with $d\tilde{x}=0.05$ to present the particle-emission results.

\end{document}